\documentclass [11pt, a4paper]{article}
\begin{document}
\begin{center}
\textbf{\large{Semi-localized axial state of a relativistic electron}}
\end{center}
\begin{center}
Tikhonenkov Igor

\textit{Department of Physics, Israel Institute of Technology, Haifa, Israel}\footnote{e-mail: itikhonen@hotmail.com}
\end{center}

\small{The new quantum state of a relativistic electron in a vacuum is described. It corresponds to an electron moving freely along a certain direction and being self-localized in a plane which is transverse to its momentum. This semi-localized state is self-consistent with respect to the interaction of an electron with its own electromagnetic field. The width of a transverse distribution is proved to be the Compton wavelength of an electron}

\vspace{0.5cm}
\normalsize
Quantum electrodynamics [1] describes an electron in a vacuum by a bispinor whose dependence upon space-time coordinates \textbf{r}, \textit{t} is reduced to a plane wave form $\exp\left(i(\mathbf{pr}-Et)/\hbar\right)$, where  \textbf{p}, \textit{E} stand for momentum and energy, $\hbar$ - Plank constant. Here I report about a solution of Dirac equation which corresponds to an electron which propagates as a free particle along some direction, say the Cartesian axis \textit{z}, but being localized in a transverse plane. The latter property results in an imaginary transverse momentum, that is $\mathbf{p} = \left(i\mathbf{p}_{\bot}, p_z \right)$, so the dispersion equation becomes
\begin{equation}
E^2/c^2-m^2 c^2 = p_z^2 -p_{\bot}^2
\end{equation}
where \textit{c} is the velocity of light in a vacuum, \textit{m} is a rest mass of an electron. It implies that a corresponding bispinor includes a factor $\exp\left(-\mathbf{p}_{\bot}\mathbf{r}_{\bot}/\hbar\right)$ or, restricting consideration to states with a zero orbital momentum, $\exp\left(-p_{\bot}r_{\bot}/\hbar\right)$. Here 
$r_{\bot}>0$ is a distance from z-axis. The positive quantity $\lambda_{\bot} =\hbar/p_{\bot}$ now means an inverse width of a localized state, which has a density distribution of mass and charge. Therefore such a particle is not a point-like one and the interaction with its own field has to be included.

In what follows first such a solution of Dirac equation for a free electron will be presented and afterwards I'll show that among these semi-localized plane waves there exist such one which retain its form if the interaction with an electromagnetic field of an electron has been taken into account.

Dirac equation in its standard representation reads:
\begin{equation}
\left(\hat{p}_0\gamma^0 -\hat{\mathbf{p}}\mbox{\boldmath${\gamma}$} \right)\Psi=mc\Psi
\end{equation}
where $\Psi$  is a bispinor, momentum operators are $\hat{p}_0=i\hbar\partial_t /c , \hat{\mathbf{p}}=-i\hbar\nabla$ , Dirac matrices:

\begin{center}$
\gamma^0=\left( 
\begin{array}{cc}
1 & 0\\
0 & -1
\end{array}
\right), \quad
\mbox{\boldmath${\gamma}$}=\left(
\begin{array}{cc}
0 & \mbox{\boldmath${\sigma}$}\\
- \mbox{\boldmath${\sigma}$} & 0
\end{array}
\right) .
$\end{center}
Cartesian components of the matrix vector $\mbox{\boldmath${\sigma}$}$ are Pauli matrices
\begin{center}$
\sigma_x = \left( 
\begin{array}{cc}
0 & 1\\
1 & 0
\end{array}
\right), \quad
\sigma_y = \left( 
\begin{array}{cc}
0 & -i\\
i & 0
\end{array}
\right), \quad
\sigma_z = \left( 
\begin{array}{cc}
1 & 0\\
0 & -1
\end{array}
\right).
$\end{center}

It is customary to use cylinder coordinates  $r_{\bot}, \alpha, z$ where the operator $\mbox{\boldmath${\sigma}$}\nabla$ has the form
\begin{center}$
\mbox{\boldmath${\sigma}$}\nabla = \left(
\begin{array}{cc}
\partial_z & e^{-i\alpha}(\partial_{r_{\bot}}-i\partial_{\alpha}/r_{\bot})\\
e^{i\alpha}(\partial_{r_{\bot}}+i\partial_{\alpha}/r_{\bot}) & -\partial_z
\end{array}
\right)
$\end{center}

The axially localized solution of (2) reads
\begin{equation}
\Psi=\psi(r_{\bot})\exp(i(k_{z}z-Et/\hbar))\left(\begin{array}{c} \mathbf{w}(\alpha)\\ \mathbf{v}(\alpha)\end{array} \right), \quad
\psi(r_{\bot}) = C\exp\left(-r_{\bot}k_{\bot}\right)/\sqrt{r_{\bot}}
\end{equation}
where $C$ is a normalizing constant, $p_z =\hbar k_{z} , p_{\bot} =\hbar k_{\bot}$. Here two-component quantities \textbf{w}, \textbf{v} depend upon an azimuth angle $\alpha$  according to
\begin{center}$
\mathbf{w}(\alpha)=\left(\begin{array}{c} w_{1}e^{-i\alpha/2}\\ w_{2}e^{i\alpha/2}\end{array}\right),\quad
\mathbf{v}(\alpha)=\left(\begin{array}{c} v_{1}e^{-i\alpha/2}\\ v_{2}e^{i\alpha/2}\end{array}\right)
$\end{center}
and
\begin{eqnarray*}
\left(\begin{array}{c} v_{1}\\ v_{2}\end{array}\right) & = & \sqrt{\frac{\epsilon -1}{\epsilon +1}} \left(\begin{array}{cc} 
\cosh(\beta/2) & i\sinh(\beta/2)\\  i\sinh(\beta/2) & -\cosh(\beta/2)
\end{array}\right) \left(\begin{array}{c} w_{1}\\ w_{2}\end{array}\right)
\end{eqnarray*}
where the rescaled energy $\epsilon = E/mc^2, \tanh(\beta/2)=k_{\bot}/k_{z}, \beta > 0$. In addition it must be $\left| w_{1}\right |^2 + \left| w_{2}\right|^2 = 1$, so up to a phase factor we may take $w_{1}=\cos(\theta/2), w_{2}=\sin(\theta/2)e^{i\varphi} (0\leq \theta \leq \pi, 0\leq \varphi \leq 2\pi)$.

The electron interacting with an electromagnetic field is described by the equation [2]:
\begin{equation}
\left(\left(\hat{p}_0 - e\Phi/c \right)\gamma^{0}-\left(\hat{\mathbf{p}}-e\mathbf{A}\right)\mbox{\boldmath${\gamma}$}\right)\Psi = mc\Psi
\end{equation}
where $\Phi, \mathbf{A}$ are scalar and vector potentials (in ratinalized mks units). They are determined by equations [3]:
\begin{equation}
\nabla^2 \Phi - \partial_{t}^2\Phi/c^2 = -\varrho_{c}/\varepsilon_0 , \quad  \nabla^2 \mathbf{A} - \partial_{t}^2\mathbf{A}/c^2 =-\mu_{0}\mathbf{j}_{c}
\end{equation}
The densities of charge and current $\varrho_{c}, \mathbf{j}_{c}$ are defined through densities of particles $n$ and of their flux \mbox{\boldmath$\nu$}  as [2]:
\begin{equation}
\varrho_{c}=en, n=\bar{\Psi}\gamma^0 \Psi,\quad \mathbf{j}_{c}=ec\mbox{\boldmath$\nu$}, \mbox{\boldmath$\nu$}= \bar{\Psi}\mbox{\boldmath$\gamma$} \Psi
\end{equation}
where Dirac conjugation $\bar{\Psi} = \Psi^{+}\gamma^0$, $\Psi^{+}$ is Hermitian conjugate of $\Psi$, \textit{e} is a charge of an electron. The requirement that $\Psi$ is the same as (3) leads to that field potentials doesn't depend upon $t$ and $z$, $\alpha$ because r.h.s. of (5) does not. Thus, using 
$\nabla^2 \mathbf{A} = \nabla (\nabla \cdot \mathbf{A})- \nabla \times (\nabla \times \mathbf{A})$, (5) may be reduced to the following ordinary differential equations:
\begin{eqnarray}
\Phi^{\prime\prime} + \Phi^{\prime}/r_{\bot} & = & -(e\tilde{n}/\varepsilon_0)\left|\psi(r_{\bot}) \right|^2 \nonumber \\
A_z^{\prime\prime} + A_z^{\prime}/r_{\bot} &=& -\mu_{0}ec \tilde{\nu}_{z} \left|\psi(r_{\bot}) \right|^2 \nonumber \\
A_{\alpha}^{\prime\prime} + A_{\alpha}^{\prime}/r_{\bot} - A_{\alpha}/r_{\bot}^2 &=& -\mu_{0}ec \tilde{\nu}_{\alpha} \left|\psi(r_{\bot}) \right|^2\\
A_{r_{\bot}}^{\prime\prime} + A_{r_{\bot}}^{\prime}/r_{\bot} - A_{r_{\bot}}/r_{\bot}^2 &=& -\mu_{0}ec\tilde{\nu}_{r_{\bot}}\left|\psi(r_{\bot}) \right|^2 \nonumber
\end{eqnarray}
where the prime designates $d/dr_{\bot}$ and 
\begin{eqnarray}
\tilde{\nu}_z & = & 2\sqrt{\frac{\epsilon -1}{\epsilon +1}}\left(\cosh(\beta/2)-\sinh(\beta/2)\sin(\theta)\sin(\varphi) \right) \nonumber \\
\tilde{\nu}_{\alpha} & = & 2\sqrt{\frac{\epsilon -1}{\epsilon +1}}\sinh(\beta/2)\cos(\theta) \nonumber \\
\tilde{\nu}_{r_{\bot}} & = &0 \\
\tilde{n} &=& 1+\frac{\epsilon -1}{\epsilon +1}\left(\cosh(\beta)-\sinh(\beta)\sin(\theta)\sin(\varphi) \right) \nonumber
\end{eqnarray}
Again wishing that the solution of (2) will obey (4) one conludes that the latter may contain  only $\Phi$, $A_z$ or $A_{\alpha}$, $A_{r_{\bot}}$ which may be cancelled by addition or substraction. Since $\Phi \neq 0$ for certain, then $A_{r_{\bot}}=A_{\alpha}=0, \tilde{\nu}_{\alpha}=0$ which gives
\begin{equation}
\cos\theta = 0
\end{equation}

To make further steps clearer I rewrite equations (4), (7) using dimentionless quantities. The unit of length is now Compton wave-length of electron $\lambda_{c} = \hbar/mc$. The rescaled distance $\rho = r_{\bot}/\lambda_{c}$, rescaled field potentials $\phi=e\Phi/mc^2 , \mathbf{a}=e\mathbf{A}/mc$, rescaled wave numbers $q_{z}=\lambda_{c}k_{z}, q_{\bot}=\lambda_{c}k_{\bot}$. The dispersion equation (1) now reads
\begin{center}$
\epsilon^2 -1=q_{z}^{2}-q_{\bot}^2
$\end{center}
so $q_z = \sqrt{\epsilon^2 -1}\cosh(\beta/2), q_{\bot} = \sqrt{\epsilon^2 -1}\sinh(\beta/2)$. The constant $C$ is defined by the condition $\int{n(\mathbf{r})d\mathbf{r}} = 1$ which makes final solutions to be normalized for one particle per volume. It gives for a radial part of $\Psi$:
\begin{equation}
\left|\psi(\rho)\right|^2=\frac{q_{\bot}}{2\pi\ell_z}\frac{\exp(-2\rho q_{\bot})}{\rho\lambda_c^3}
\end{equation}
where $\ell_z$ is a some macroscopic quantization length in $\lambda_c$ units. Insertion of $\Psi$ of the form (3),(10) into (4),(7) gives the following set of equations:

\begin{eqnarray*}
(\epsilon -\phi(\rho)-1) \left(\begin{array}{c} w_1 \\ w_2 \end{array}\right) & = &\frac{1}{(\epsilon +1)}
\left(\begin{array}{cc} q_z - a_z (\rho) & iq_{\bot} \\ iq_{\bot} & a_z(\rho) - q_z \end{array} \right)
\left(\begin{array}{cc} q_z & iq_{\bot} \\ iq_{\bot} & -q_z \end{array} \right)
\left(\begin{array}{c} w_1 \\ w_2 \end{array} \right)
\end{eqnarray*}
\begin{eqnarray*}
\left(\begin{array}{cc} q_z - a_z (\rho) & iq_{\bot} \\ iq_{\bot} & a_z(\rho) - q_z \end{array} \right)
\left(\begin{array}{c} w_1 \\ w_2 \end{array}\right) & = &\frac{\epsilon-\phi(\rho)+1}{\epsilon+1}
\left(\begin{array}{cc} q_z & iq_{\bot} \\ iq_{\bot} & -q_z \end{array} \right)
\left(\begin{array}{c} w_1 \\ w_2 \end{array}\right)
\end{eqnarray*}
It is easy to see that it holds when the fields are abcent, that is $\phi=0, \mathbf{a}=0$. Reducing those parts we have
\begin{eqnarray}
\left(\begin{array}{cc} 
(\epsilon+1)\phi(\rho)-q_z a_z (\rho) & -iq_{\bot}a_{z}(\rho) \\ 
iq_{\bot}a_{z}(\rho) & (\epsilon+1)\phi(\rho)-q_z a_z (\rho)  
\end{array} \right) \left(\begin{array}{c} w_1 \\ w_2 \end{array}\right) & = & 0  \nonumber \\
\left(\begin{array}{cc}
(\epsilon-1)\phi(\rho)-q_z a_z (\rho) & iq_{\bot}a_{z}(\rho) \\ 
-iq_{\bot}a_{z}(\rho) & (\epsilon-1)\phi(\rho)-q_z a_z (\rho)  
\end{array} \right) \left(\begin{array}{c} w_1 \\ w_2 \end{array}\right) & = & 0
\end{eqnarray}

The equations (6) for potentials are resolved as
\begin{eqnarray}
\phi(\rho) & = &- \frac{\alpha_f}{\ell_z}\tilde{n}\int_0^{\rho}{\frac{dx}{x}(1-\exp(-2q_{\bot}x))} \nonumber \\
a_z(\rho) & = &- \frac{\alpha_f}{\ell_z}\tilde{\nu_z}\int_0^{\rho}{\frac{dx}{x}(1-\exp(-2q_{\bot}x))}
\end{eqnarray}
where $\alpha_f  = e^2 / (4\pi\varepsilon_0 \hbar c)$ is the fine structure constant.
Next step is to detect whether exist such $\beta > 0,\theta,\varphi$ that the equations (11) have at least one nontrivial solution e.g. $ \left| w_1 \right|^2 + \left| w_2 \right|^2 \neq 0$.  The equations (11) are consistent if
\begin{equation}
\epsilon\phi = q_z a_z ,\quad \phi^2 = q_{\bot}^2 a_z^2 
\end{equation}
where the dependence upon $\rho$ is omitted for brevity. If (13) is true then it follows from (11) that
\begin{equation}
\phi w_1 - i q_{\bot}a_z w_2 =0, \quad iq_{\bot}a_z w_1 +\phi w_2 =0
\end{equation}
The equations (9), (13), (14) hold together if
\begin{equation}
\theta = \pi/2, \quad \varphi = -\pi/2, \quad \tanh(\beta/2) = 1/\epsilon
\end{equation}
Thus there is only one semilocalized plane wave (3) which retains its form under presence of electron's own field. For this state $q_{\bot} =1$ which means that the transverse width of the distribution of mass (and charge) in a transverse plane is  $\lambda_c$.

\vspace{1cm}

\textbf{Referencies.} 
\begin{enumerate}
\item B. Berestetski, E. Liphshits, L. Pitaevski, Quantum electrodynamics, \S 23,Moscow, Nauka,1979 (\textit{in Russian}).
\item Ibidem, \S 32.
\item J. Jackson, Classical electrodynamics, \S 6.4, John Wiley \& Sons, 1962
\end{enumerate}
\end{document}